\documentclass{aip-cp}

\usepackage[numbers,elide]{natbib}
\usepackage{fancybox}
\usepackage{shortvrb}
\MakeShortVerb\|

\usepackage{listings}
\usepackage[usenames]{color}
\definecolor{codecolor}{gray}{.9}
\definecolor{rlcolor}{cmyk}{0,1,0,0}
\begin{document}

\title{Ab initio Simulations of Superionic H$_2$O, H$_2$O$_2$, and H$_9$O$_4$ Compounds}

\author[aff1,aff2]{Burkhard Militzer\corref{cor1}}
\author[aff1,aff3]{Shuai Zhang}
\affil[aff1]{Department of Earth and Planetary Science, University of California, Berkeley, CA 94720, USA}
\affil[aff2]{Department of Astronomy, University of California, Berkeley, CA 94720, USA}
\affil[aff3]{Lawrence Livermore National Laboratory, Livermore, California 94550, USA}
\corresp[cor1]{Corresponding author: militzer@berkeley.edu}

\maketitle

\newcommand{\SSS}{{\rm SS}}
\newcommand{\PP}{{\rm p}}
\newcommand{\gcc}{{g$\,$cm$^{-3}$}}
\newcommand{\etal}{{\it et al.}}
\newcommand{\figurewidth}{84mm}
\newcommand{\smallfigurewidth}{81mm}
\newcommand{\smallerfigurewidth}{78mm}

\begin{abstract}
  Using density functional molecular dynamics simulations, we study
  the behavior of different hydrogen-oxygen compounds at megabar
  pressures and several thousands of degrees Kelvin where water has
  been predicted to occur in superionic form. When we study the close
  packed hcp and dhcp structures of superionic water, we find that
  they have comparable Gibbs free energies to the fcc structure that
  we predicted previously [Phys.  Rev.  Lett., 110 (2013) 151102].
  Then we present a comprehensive comparison of different superionic
  water candidate structures with $P2_1$, $P2_1/c$, $P3_121$, $Pcca$,
  $C2/m$, and $Pa\bar 3$ symmetry that are based on published
  ground-state structures. We find that the $P2_1$ and $P2_1/c$
  structures transform into a different superionic structure with
  $P2_1/c$ symmetry, which at 4000 K has a lower Gibbs free energy than fcc for
  pressures higher than 22.8 $\pm$ 0.5 Mbar. This novel structure may
  also be obtained by distorting a hcp supercell. Finally we show that
  H$_2$O$_2$ and H$_9$O$_4$ structures will also assume a superionic
  state at elevated temperatures. Based on Gibbs free energy
  calculations at 5000 K, we predict that superionic water decompose
  into H$_2$O$_2$ and H$_9$O$_4$ at 68.7 $\pm$ 0.5 Mbar.
\end{abstract}


\section{INTRODUCTION}

Water is one of the most prevalent substances in the solar
system~\cite{hubbard_planets}. Since it existed in solid form beyond
the ice line around the early sun, it was incorporated very
efficiently into the four forming giant planets. These planets thus
grew much more rapidly than the terrestrial planets in our solar
system and reached a critical size early enough to accret a
substantial amount of gas before that was driven away by the solar
wind.

Characterizing water at the condition of giant planet interiors of
megabar pressure and thousands of degrees Kelvin is therefore crucial
for our understanding the interior structure and evolution of these
planets.  Uranus and Neptune have unusual, non-dipolar magnetic fields
that is assumed to be generated in the ice
layer~\cite{Ness1986,Stanley2004}.  Water may occur there in
superionic form where the oxygen atoms are held in place like atoms in
crystal while the much smaller hydrogen atoms diffuse throughout the
oxygen sub-lattice like a fluid. Such a state would have a high ionic
conductivity and contribute to the magnetic field generation.

While {\it ab initio} simulations consistently predicted water to
assume a superionic
state~\cite{cavazzoni,goldman2005,mattson2006,french-prb-09,WilsonWongMilitzer2013,caracas2016},
experimental confirmation that such a state exists has proven to be a
challenge~\cite{goncharov2009}. Static compression experiments in
diamond anvil cells have reached pressures up to 2.1
Mbar~\cite{hemley-ice-1987,Goncharov1996,Loubeyre1999}. Shock
wave experiments~\cite{lee2006_short,KnudsonWater2012} have also
reached higher pressures but they heat the sample significantly so
that it melts for the highest pressures. However, dynamic ramp
compression techniques~\cite{WangRamp2013} are expected to reach high
pressures at comparatively low temperatures, where so far only
theoretical methods have predicted the state of water.

Over the course of the last seven years, much theoretical effort has
been put into characterizing the state of water ice at megabar
pressures.  Starting with Ref.~\cite{MilitzerWilson2010}, a series of
structures have been predicted for the
ground-state~\cite{McMahon2011,Ji2011,WangMa2011,Hermann2011,H4O}. A
H$_4$O stoichoimetry was predicted to become stable at 14
Mbar~\cite{H4O}.  Soon after, Pickard \etal~\cite{Pickard2013}
predicted H$_2$O ice would cease to exist at approximately 50 Mbar and
decompose into H$_2$O$_2$ and an hydrogen-rich structure like
H$_9$O$_4$. As we will demonstrate in this article, a similar
decomposition occurs in superionic water.  H$_2$O$_2$ and H$_9$O$_4$
also become superionic at elevated temperatures. Our {\it ab initio}
Gibbs free energy calculations show the decomposition of superionic
H$_2$O is favored for pressures above 68.7 $\pm$ 0.5 Mbar.

We previously predicted a transition from a body-centered cubic (bcc) to
face-centered cubic (fcc) oxygen sub-lattice in superionic water at
approximately 1 Mbar~\cite{WilsonWongMilitzer2013}. In this paper, we
show with {\it ab initio} simulations that other close packed
structures have Gibbs free energies that are very similar to that of
fcc and we can therefore no longer predict with certainty which close
packed structure superionic water will assume up to $\sim$22.8 $\pm$ 0.5
Mbar.  Consistent with the findings in Ref.~\cite{Princeton}, we
predict superionic water assume a novel structure with $P2_1/c$
symmetry that is not close packed but more dense than fcc at the same
$P$-$T$ conditions~\cite{APS}. This transformation is accommodated by
changes in the oxygen sub-lattice.

\section{DENSITY FUNCTIONAL MOLECULAR DYNAMICS SIMULATIONS AND GIBBS FREE ENERGY CALCULATIONS}

All {\it ab initio} simulations were based on density functional
molecular dynamics (DFT-MD) and performed with the VASP
code~\cite{vasp}. We used pseudopotentials of the projector-augmented
wave type~\cite{paw} with core radii of 1.1 and 0.8 Bohr for the O and
H atoms respectively. For most calculations, we used the
exchange-correlation functional of Perdew, Burke and
Ernzerhof~\cite{PBE} but we employed the local density approximation
(LDA) for comparison. A cutoff energy of 900~eV for the plane wave
expansion of the wavefunctions was used throughout. The Brillioun zone
was sampled with $2 \times 2 \times 2$ Monkhorst-Pack $k$-point
grids~\cite{MP76}. $4 \times 4 \times 4$ grids had been
tested~\cite{WilsonWongMilitzer2013} and yielded consistent results.
The occupation of electronic states are taken to be a Fermi-Dirac
distribution set at the temperature of the ions~\cite{mermin}. The
simulation time ranged between 1.0 and 5.0 ps.  A time step of 0.2 fs
was used.

An initial configuration for a particular structure is obtained by
starting from the ground state and gradually increasing the
temperature during the DFT-MD simulation until the hydrogen atoms are
mobile and equilibration is reached. Positions and velocities of
equilibrated superionic structures were recycled to initialize
simulations at other densities and temperatures after adjusting the
cell parameters or velocities, respectively. We found that all ice
structures transform into a superionic state at megabar pressures.
Depending on the structure, the solid-to-superionic transition may be
accompanied by re-arrangements of the oxygen sub-lattice that needs to
be analyzed for every structure and density.

All simulations were performed in supercells with between 48 and 108
molecules. We used our algorithm~\cite{supercells} to construct
compact supercells of cubic or nearly cubic shape by starting from the
primitive cell vectors $\vec{a}_{\PP}$, $\vec{b}_{\PP}$, and $\vec{c}_{\PP}$:
\begin{equation}
\nonumber
\vec{a} = i_a \vec{a}_{\PP} + j_a \vec{b}_{\PP} + k_a \vec{c}_{\PP}\;\;\;\;,\;\;\;\;
\vec{b} = i_b \vec{a}_{\PP} + j_b \vec{b}_{\PP} + k_b \vec{c}_{\PP}\;\;\;\;,\;\;\;\;
\vec{c} = i_c \vec{a}_{\PP} + j_c \vec{b}_{\PP} + k_c \vec{c}_{\PP}\;\;\;\;.
\end{equation}
The integer values, $(i_a~j_a~k_a,~i_b~j_b~k_b,~i_c~j_c~k_c)$, will be
specified in each case.

While some structures like fcc have no
adjustable cell parameters, other less symmetric crystal structures
may have up to five free parameters, the $b/a$ and $c/a$ ratios as
well as the three angles, for a given volume and temperature. Most
simply one can derive these parameters with ground-state structural
relaxation.  Since all superionic properties are neglected in such
optimizations, the results need to be confirmed with finite
temperature DFT-MD simulations. If there is just one free parameter,
such as the $c/a$ ratio in the hcp structure, one can perform a series
of constant-volume simulations for different $c/a$ ratios and
determine the optimal value by fitting a linear stress-strain
relationship, $ \sigma_{zz} - (\sigma_{xx}+\sigma_{yy}+\sigma_{zz})/3
= A (c/a) +B$, where $\sigma_{ij}$ are the time-averaged components of
the stress tensor. The optimal c/a ratio is given by $-B/A$ when the
system is under hydrostatic conditions.

For systems with more free cell parameters, it may be more efficient
to perform constant-pressure, flexible cell
simulations~\cite{Hernandez2001}. All free parameters will then
fluctuate around a mean value that can be estimated by taking a simple
average. Since the fluctations may be quite large, one can introduce a
bias into the average. By monitoring the stress value during the MD
simulations, one may only consider cell parameter value for the
average if the instantaneous stress coincides with the target stress.
Alternatively, one may use the the stress-strain pairs from the MD
trajectory to fit a linear relationship $\sigma_{ij}-\sigma_{ij}^{\rm
  target} = A \epsilon_{ij}+B$. The optimal strain value is again
given by $-B/A$.  We have tested all these three methods and found
them to have comparable accuracy, which is primarily controlled by the
length of the MD trajectory.

We derive the {\it ab initio} Gibbs free energies with a
thermodynamic integration (TDI)
technique~\cite{Wijs1998,Morales2009,WilsonMilitzer2010,WilsonMilitzer2012,WilsonMilitzer2012b,Militzer2013,Wahl2013,Gonzalez2015,Soubiran2015,Soubiran2016}
that we adapted to superionic systems in
Ref.~\cite{WilsonWongMilitzer2013}. In this scheme, the
difference in Helmholtz free energy between a DFT system and a system
governed by classical forces is computed from,
\begin{equation}
F_{\rm DFT} - F_{\rm cl} = \int_0^1 d\lambda \; \left< V_{\rm KS} - V_{\rm cl} \right>_\lambda.
\label{tdint}
\end{equation}
The angle brackets represent an average over trajectories governed by
forces that are derived from a hybrid potential energy function,
$V_\lambda=V_{\rm cl} + \lambda (V_{\rm KS}-V_{\rm cl})$. $V_{\rm cl}$ is
the potential energy of the classical system and $V_{\rm KS}$ is the
Kohn-Sham energy. We typically perform five independent simulations
with $\lambda$ value equally spaced between 0 and 1.

In order to maximize the efficiency of evaluating the integral in
Eq.~(\ref{tdint}), we construct a new set of classical forces for each
density and temperature. Our classical reference system consists of a
pair potentials between each pair of atomic species combined with a
harmonic Einstein potential for every oxygen atom. The harmonic force
constants are derived first from the mean square displacements in a
constant-volume simulation. The residual forces are fitted to O-O,
O-H, and H-H pair potentials~\cite{forcematching} that we present with
spline functions~\cite{Wahl2015}.

The free energy of the classical reference system, $F_{\rm cl}$, is
obtained with classical Monte Carlo simulations where we again take
advantage of the TDI technique to gradually turn off all pair forces.
The free energy of an Einstein crystal of oxygen atoms and that of a
gas of noninteracting hydrogen atoms is known analytically.

\section{COMPARISON OF DIFFERENT H$_2$O STRUCTURES}

In Ref.~\cite{WilsonWongMilitzer2013}, we predicted superionic
water to assume a fcc structure at megabar pressures
while earlier {\it ab initio} simulations had assumed a
bcc oxygen sub-lattice. We begin the discussion in
this section by comparing the Gibbs free energy of fcc with that of
other close packed structures. In particular we will consider a
hexagonal close packed (hcp) structure with an ABAB layering and
double hexagonal close packed (dhcp) structure with an ABAC
layering~\cite{MaSodium2008}.

\begin{figure}[htb]
\includegraphics[width=140mm]{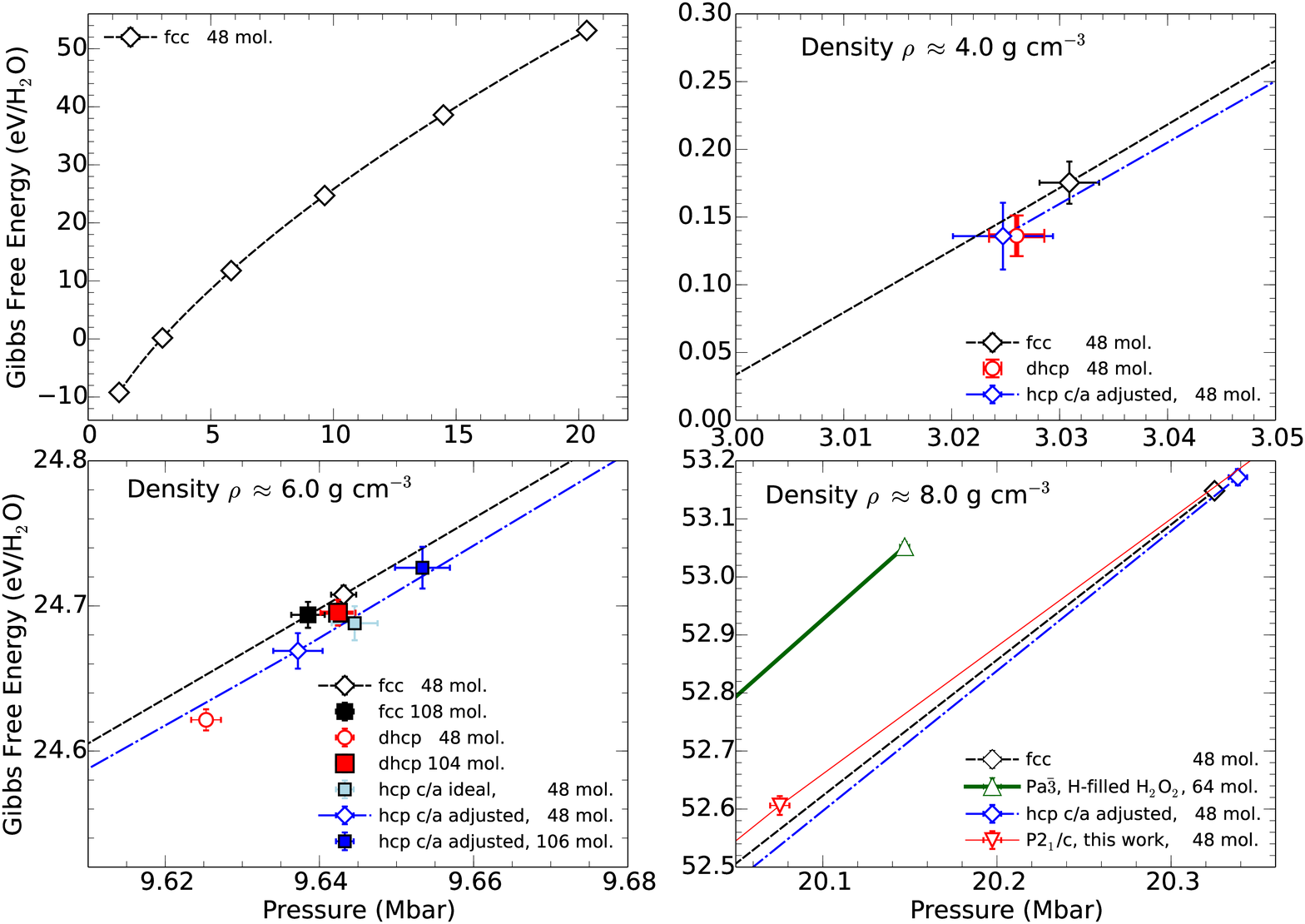}
\caption{Gibbs free energy as a function of pressure at
  4000 K.  While the left upper shows results for the fcc structure
  over a large pressure range, the remaining diagrams compare
  different structure at a particular density. The number of H$_2$O
  molecules in the supercell is indicated in the caption.}
\label{GP}
\end{figure}

We performed Gibbs free energy calculations for fcc, hcp, and dhcp
structures for a number of densities and supercell sizes at a
temperature of 4000 K. Figure~\ref{GP} compares our results for 4.0,
6.0, and 8.0 \gcc. When we performed calculation for dhcp at 6.0 \gcc~
using a (4 2 0, 0 -3 0, 0 0 -1) supercell with 48 molecules, we
obtained a Gibbs free energy that was slightly below that of fcc. A
further analysis revealed, however, that this deviation was caused by
finite size effects. Our simulation of dhcp using a significantly
larger (3 -1 1, -3 -4 0, -2 1 1) supercell with 104 molecules yielded
a Gibbs free energy within the error bars of the fcc result.
Conversely, fcc simulations with 48 and 108 molecules using (3 1 -1, 3
-3 -1, 0 -2 4) and cubic (3 3 -3, 3 -3 3, 3 -3 -3) supercells gave
very similar Gibbs free energies. For the hcp structure, we performed
simulations with 48 and 106 molecules in (4 2 0, 0 -3 0, 0 0 -2) and
(4 0 1, 3 4 -1, 0 -3 -2) supercells. Again we find that the largest
simulations agree with the fcc Gibbs free energy. For 4.0 and 8.0
\gcc~ results are also very similar. While we favored an fcc structure
in our earlier publication~\cite{WilsonWongMilitzer2013}, we are not
able to predict with certainty which close packed structure will be
realized at megabar pressures. The magnitude of the statistical and
finite size uncertainties in our current Gibbs free energy calculation
make the distiction between different close packed superionic
structure very challenging. It is conceivable that the different
structures may be realized or co-exist depending on the specific
pressure, temperature, and type of hydrogen isotopes.

Hcp and dhcp structures have one free parameter that needs to be
adjusted, the $c/a$ ratio. At 4.0, 6.0, and 8.0 \gcc, we performed
fixed-cell simulations of hcp and dhcp for four different $c/a$ ratios
at 4000 K. We determined an optimized c/a value from a linear fit to
the stress-strain relation. For dhcp, we did not identify a
significant deviation from experimental value of 1.007$\,\times\,
(c/a)_{\rm ideal}$~\cite{MaSodium2008}, where the ideal c/a value is
given by $\sqrt{8/3}$. For hcp in the density interval from 4.0 to 8.0
\gcc, the optimized $c/a$ ratio increased from 1.005 to 1.013 times
the ideal value. However, as Fig.~\ref{GP} shows, the optimization of
the $c/a$ ratio did not lead to any significant lowering of the
Gibbs free energy compared to simulation with the ideal value.

After considering the close packed structures, we now focus on
different ground-state structures in order to test whether any of them
leads to superionic structures that are structurally stable and, more
importantly, have a lower Gibbs free energy than fcc. A hexagonal
structure with $P3_121$ symmetry has been proposed by Pickard
\etal~\cite{Pickard2013} to be the ground-state structure in pressure
interval from 8 to 14 Mbar. We constructed a (2~2~0,~1~-1~0,~0~0~-1)
supercell with 48 molecules.  After the superionic regime had been
reached in DFT-MD simulations with increasing temperature, we observed
that the oxygen sub-lattice spontaneously re-arranged from a
$P3_121$ symmetry to an fcc structure. As expected, subsequent Gibbs
free energy calculations for this structure gave the same results as
our fcc calculation within the statistical error bars. Since the
re-arrangement occurred in simulations at 6.0 and 8.0 \gcc, we have no
reason to study the $P3_121$ structure any further. In fact, this
result provides additional support for the hypothesis that fcc
structure is one of the most stable structure under these conditions.

For the pressure range from 14 to 19 Mbar, Pickard
\etal~\cite{Pickard2013} proposed an orthorhombic structure with
$Pcca$ symmetry. We performed DFT-MD simulations in
(2~0~0,~0~2~0,~0~0~1) and (2~0~0,~0~3~0,~0~0~1) supercells with 48 and
72 molecules, respectively. In both cases, the oxygen sub-lattice
changed spontaneously from a $Pcca$ symmetry to an hcp structure
during constant-volume simulations. Thus the $Pcca$ structure does not
need to be considered further and it confirms that hcp needs to be
considered as an alternative, close-packed candidate structure.

\begin{figure}[htb]
\includegraphics[width=\smallfigurewidth]{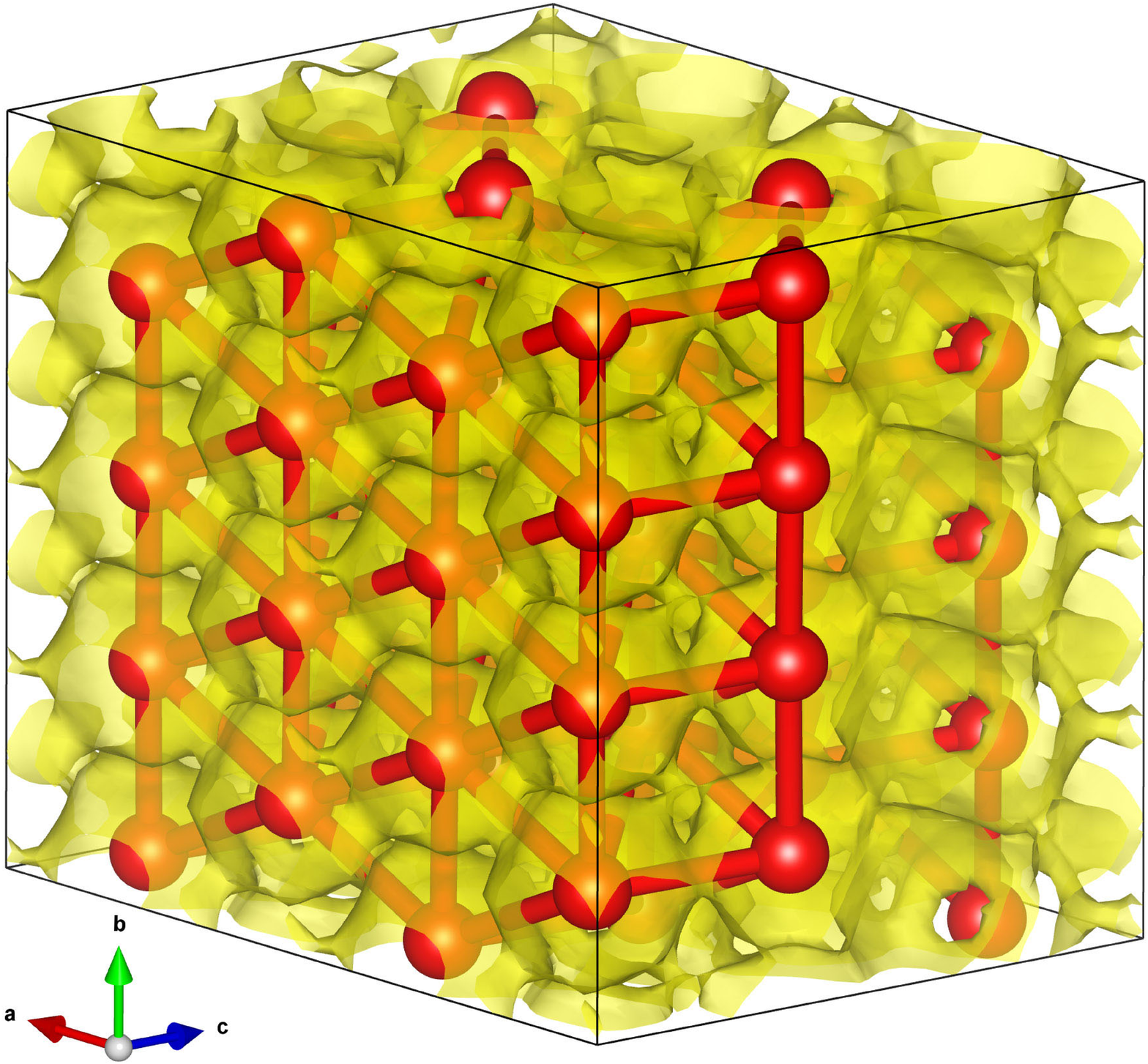}
\includegraphics[width=\smallfigurewidth]{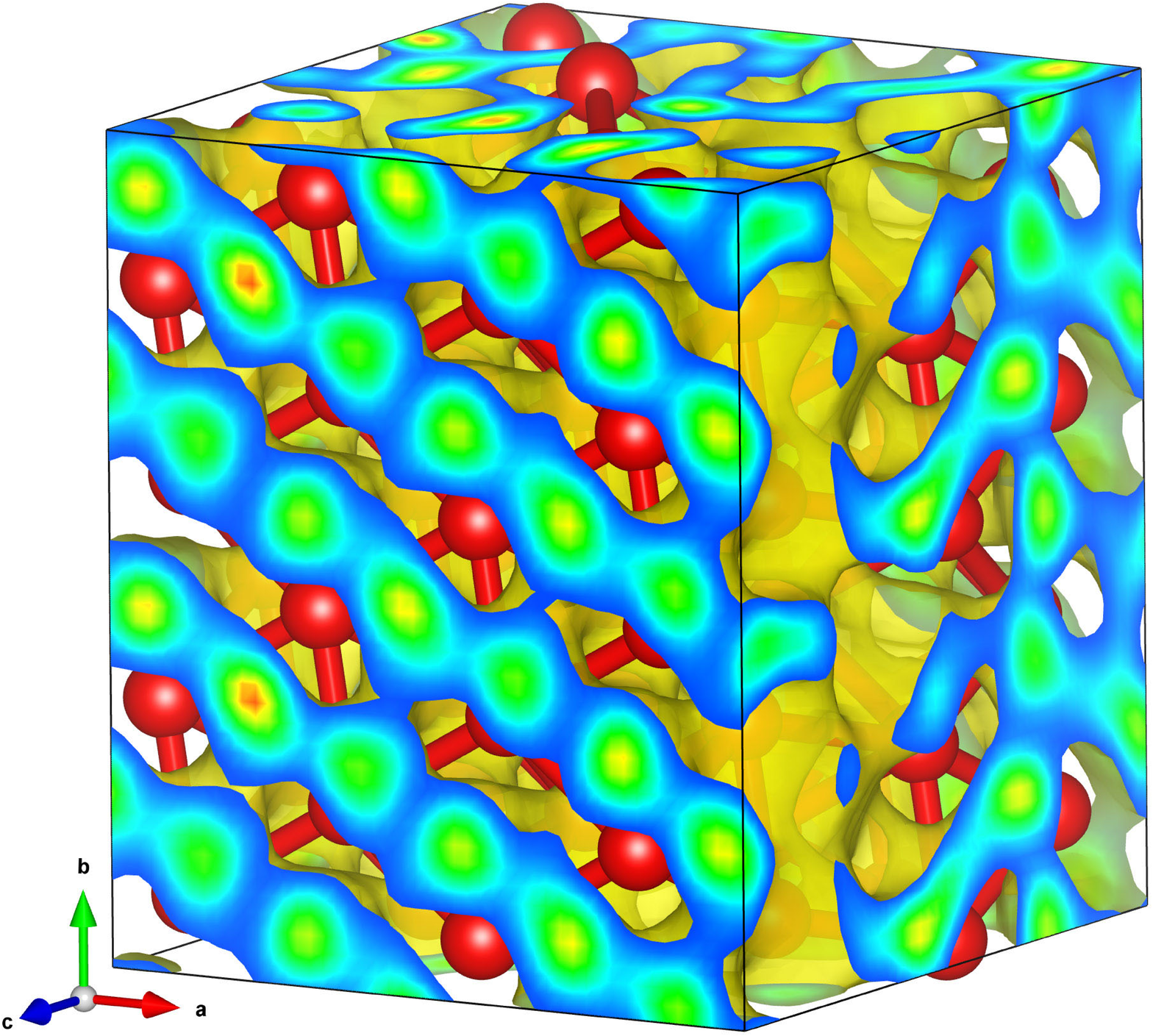}
\caption{Isosurfaces of the density of hydrogen atoms
  are compared for hcp (left) and $P2_1/c$ (right panel). The
  coloring within the isosurfaces has been omitted in the upper panel for
  clarity. The spheres denote the oxygen atoms.}
\label{hcp_P21c}
\end{figure}

McMahon predicted that water ice assumes a metallic structure with
$C2/m$ symmetry at 56 Mbar~\cite{McMahon2011}. We constructed (2 -1
0, -1 -3 -1, 0 -1 2) supercells with 64 molecules and heated them to a
superionic state. We determined the Gibbs free energy for densities
of 11.0, 11.5, and 11.9 \gcc~but found them to be consistently higher
than that of close packed structures.

\begin{table}[htb]
\caption{Lattice parameters of the monoclinic primitive 
  cell derived from simulations of our $P2_1/c$ structure 
  at different densities and temperatures. The last three columns
  specify the fractional coordinates of the oxygen atom in 
  Wyckoff position $e$ with multiplicity 4. $\dagger$ labels cell 
  parameters that we obtained by extrapolation in pressure before 
  the oxygen positions were obtained by averaging over MD trajectories.}
\label{P21c}
\begin{tabular} { c c c l l l l c c c }
\hline
$T$ (K) & $P$ (Mbar)& $\rho$~(\gcc)& $a$ (\AA)  & $b$ (\AA)   & $c$ (\AA)   &    $\beta (^\circ)$ &  $x$      & $y$      & $z$\\
\hline
5000 & 74.88  &  13.89 & 2.380 &  2.542 &  2.946  &  28.91 &  0.3516 & 0.3653 & -0.08432\\
5000 & 69.87  &  13.47 & 2.404 &  2.571 &  2.976  &  28.88 &  0.3538 & 0.3644 & -0.08483\\
5000 & 59.92  &  12.59 & 2.458 &  2.633 &  3.050  &  28.78 &  0.3486 & 0.3638 & -0.08094\\
5000 & 49.91  &  11.63 & 2.523 &  2.709 &  3.132  &  28.73 &  0.3460 & 0.3638 & -0.07871\\
5000 & 39.91  &  10.57 & 2.608 &  2.801 &  3.240  &  28.58 &  0.3417 & 0.3640 & -0.07602\\
4000 & 27.13  &  9.040 & 2.743 &  2.959 &  3.411 &   28.56 &  0.3384 & 0.3638 & -0.07367\\
4000 & 23.54  &  8.529 & 2.802 &  3.011 &  3.467 &   28.66 &  0.3293 & 0.3602 & -0.06637\\
4000 & 17.13  &  7.503 & 2.921$^\dagger$ &  3.151$^\dagger$ &  3.627$^\dagger$ &   28.53$^\dagger$ &  0.3086 & 0.3574 & -0.04738\\
\hline
\end{tabular}
\end{table}

This leaves us with two more structures to consider. Before the work
by Pickard \etal~\cite{Pickard2013}, a monoclinic structure with
$P2_1$ symmetry had been proposed by several
authors~\cite{McMahon2011,Ji2011,WangMa2011,Hermann2011} for the
pressure interval from 11.7 to 19.6 Mbar. We performed superionic
DFT-MD simulations in (3~0~0,~0~2~0,~0~0~2) and
(3~1~0,~-2~2~0,~-1~0~-2) supercells with 48 and 64 molecules,
respectively. We found that the oxygen sub-lattice re-arranges to a new
monoclinic structure with $P2_1/c$ symmetry with just four oxygen
atoms per unit cell.  Table~\ref{P21c} provides the structural
parameters for different conditions. The cell parameters were
determined first with constant-pressure simulations. The oxygen
position derived by averaging over trajectories obtained in subsequent
constant-volume simulations. Below 20 Mbar, the $P2_1/c$ structure
exhibited an instability in constant-pressure simulations. Since we
need cell parameters for Gibbs free energy calculations under these
conditions, we extrapolated the cell parameters that we obtained from
simulation at higher pressure.

Before we analyze the properties of our $P2_1/c$ structure in more
detail, we discuss the monoclinic $P2_1/c$ structure with eight
molecules per unit cell that was predicted by Ji \etal~\cite{Ji2011}
to be the ground state in the pressure range from 19.6 to 50 Mbar. We
constructed a triclinic (2~0~1,~0~2~1,~0~-1~1) and a monoclinic
(2~0~1,~0~2~0,~0~0~-2) supercell with 48 and 64 molecules,
respectively. When the simulations reached a superionic state, the
oxygen sub-lattice re-arranged itself to the same $P2_1/c$ structure
that we had obtained earlier by heating the $P2_1$ supercells.  The
resulting structural parameters were indistinguishable from those in
Tab.~\ref{P21c}.

When we analyzed the average density of hydrogen atoms in
Fig.~\ref{hcp_P21c}, we found it to be very non-uniform. This
distinguishes the $P2_1/c$ phase from close packed superionic
structures. This finding is consistent with diffusion rates being
anisotropic in this structure~\cite{Princeton}.

\begin{figure}[htb]
\includegraphics[width=\smallfigurewidth]{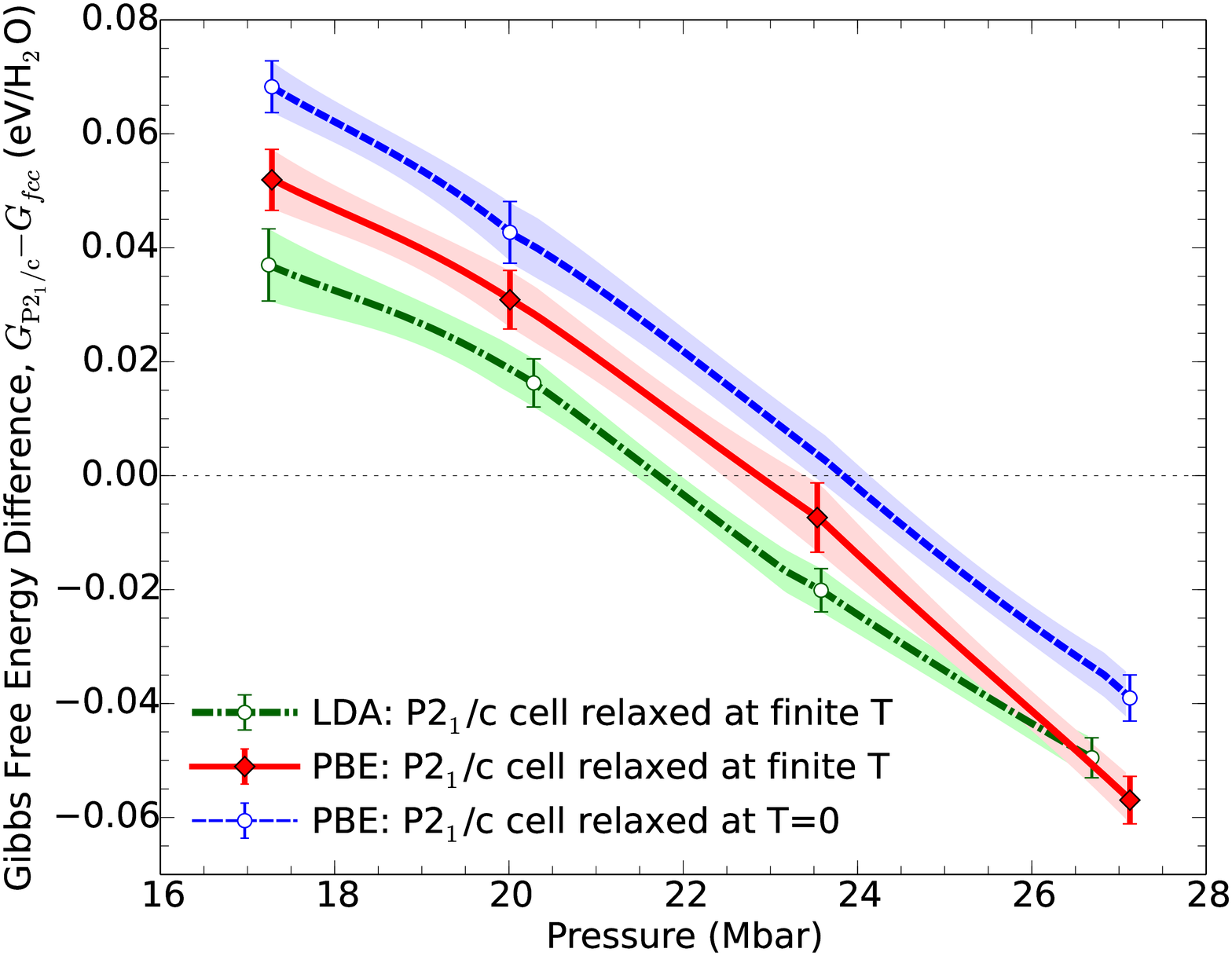}
\includegraphics[width=\smallfigurewidth]{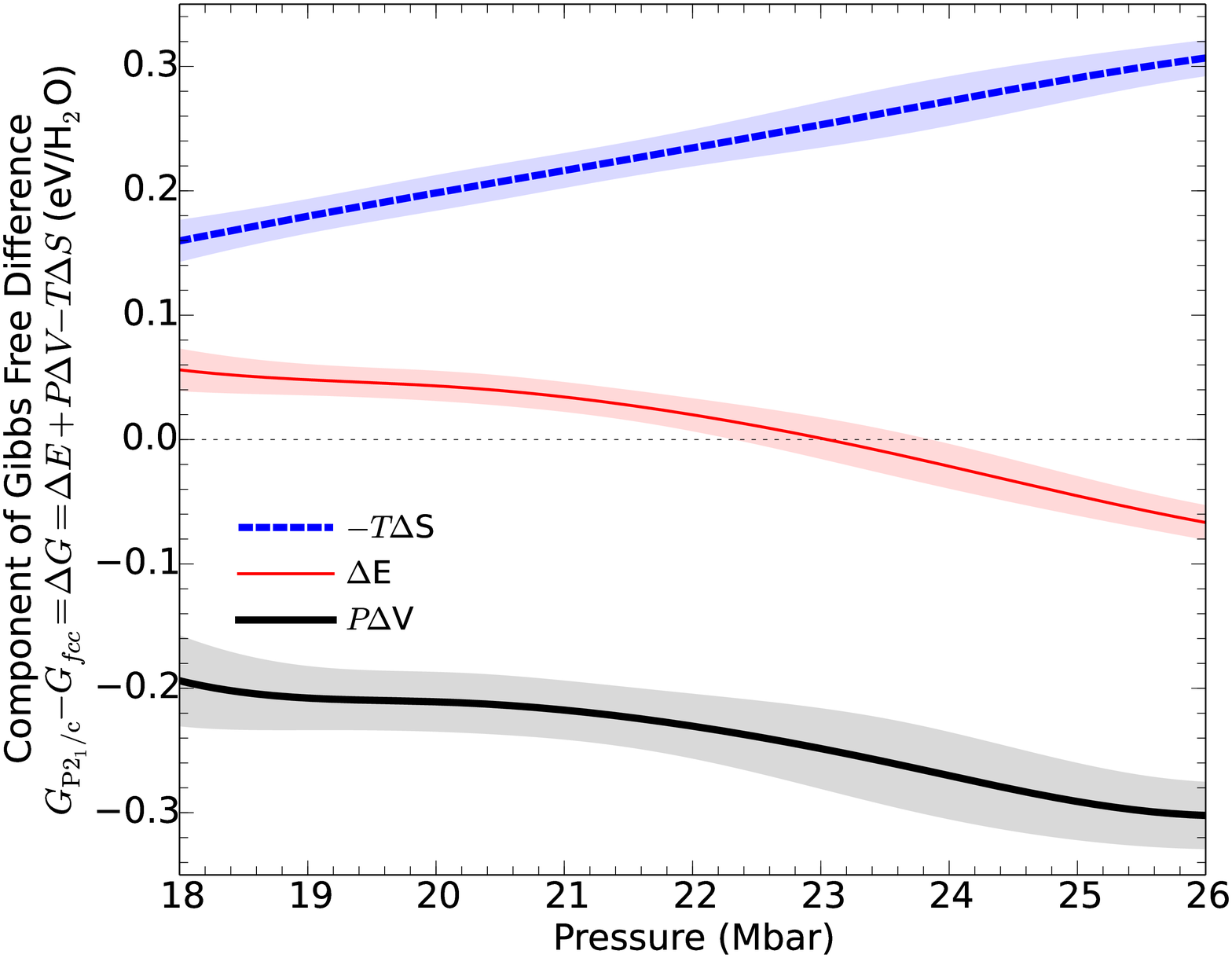}
\caption{Gibbs free energy difference between the $P2_1/c$ and fcc
  structures at 4000 K. On the left, the red and the blue curves
  correspond to two calculations of the total Gibbs free energy
  difference using the PBE functional.  For the blue curve, the
  $P2_1/c$ lattice parameters were taken from ground-state
  calculations, while for the red curve, they have been optimized with
  constant-pressure simulations. Based on the latter, more accurate
  determination, the transition from the fcc to $P2_1/c$ structure
  predicted to occur at 22.8 $\pm$ 0.5 Mbar. From comparison, a
  calculation with the LDA functional yields lower transition pressure
  of 21.7 $\pm$ 0.3 Mbar (green curve).The shaded regions indicate the
  statistical uncertainties. On the right, various components of the
  Gibbs free energy difference are displayed.}
\label{fcc_P21c}
\end{figure}

\begin{figure}[htb]
\includegraphics[width=15cm]{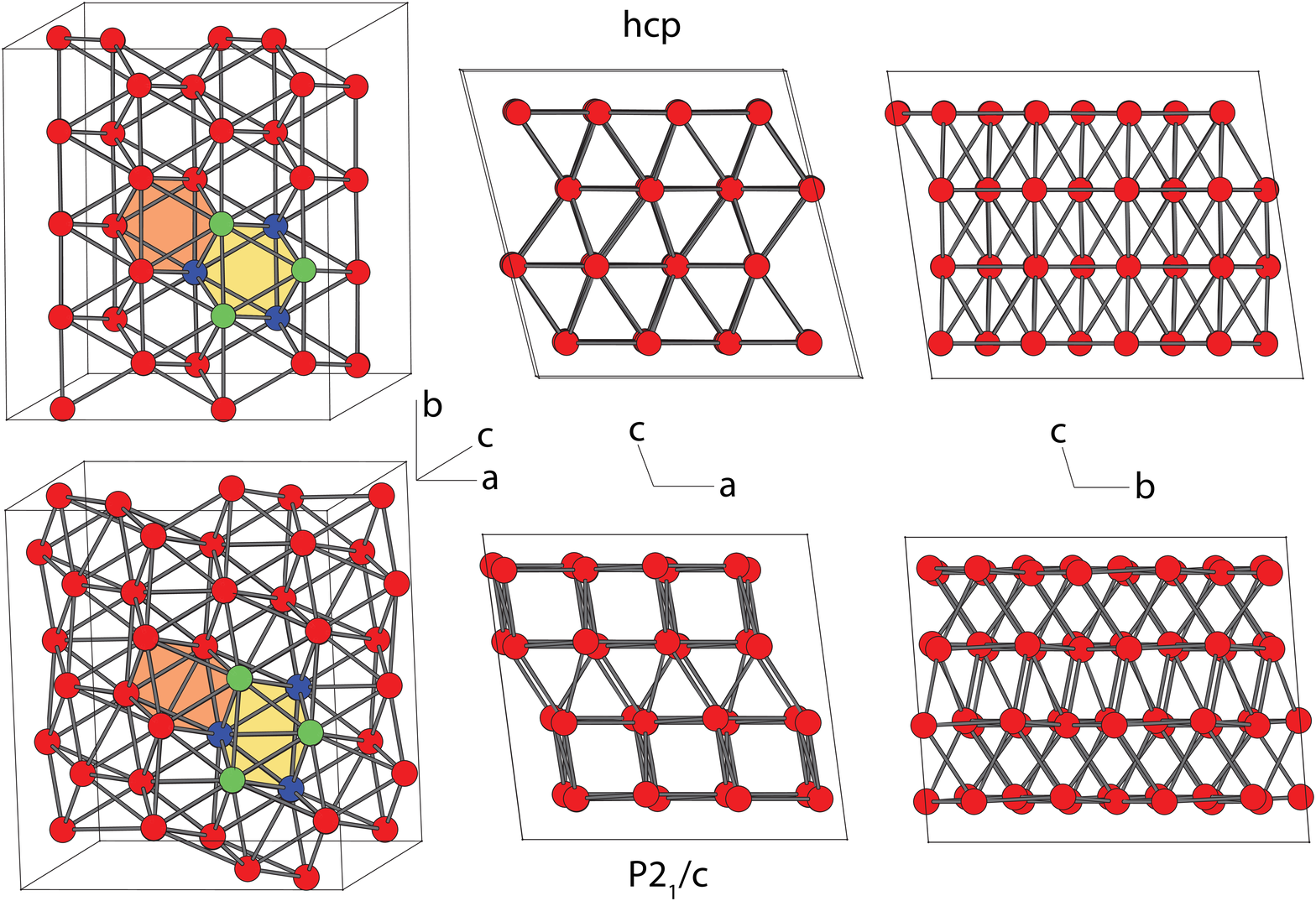}
\caption{Comparison of the hcp (upper row) and novel
  $P2_1/c$ (lower row) structures. The spheres denote the positions of
  the 64 oxygen atoms in the supercell that were derived by averaging
  over entire MD trajectories.  The lattice vectors of the supercell
  are shown schematically. On the left, two hexagons are shaded and
  three atoms in the first and second layers are colored differently
  in order to illustrate the hcp-to-$P2_1/c$ transformation that is
  facilitated by the sliding of $a$-$b$ planes.  While the diagram in
  the upper center shows the characteristic hexagonal pattern of a
  close packed structure, the corresponding diagram in the lower
  center shows stacked layers composed of nearly perfect squares and
  triangles for $P2_1/c$. On the right, the view along the $b$
  direction also illustrates modest differences.}
\label{combined_hcp_P21c}
\end{figure}

We also lowered the pressure step by step in our constant-pressure
flexible-cell simulations of the $P2_1/c$ structure in order to
identify a transformation path to a close packed structure. At 14, 17,
and 20 Mbar, the $P2_1/c$ structure transformed into a nearly perfect
hcp structure as is illustrated in Fig.~\ref{combined_hcp_P21c}. The
relaxation to a perfect hcp crystal is prevented by our choice of
supercell. This becomes apparent when the stacking of layers is viewed
along the $b$ direction. Nevertheless, the resulting structure is
sufficiently close to hcp so that we can propose a transformation path
the we illustrate in Fig.~\ref{combined_hcp_P21c}. The hcp-to-$P2_1/c$
transformation appears to be facilitated by a shift of the $a$-$b$
planes. While a view of the hcp structure along the $c$ direction
reveals perfect hexagons as a results of the ABAB layering, these
hexagons appear distorted in the $P2_1/c$ crystal as a result of the
layer shift.  In our experience, these distorted hexagons provide the
most straightforward way to identify the $P2_1/c$ structure. While in
hcp, or any other close packed system, every atom has twelve
equally-distant nearest neighbors. In the $P2_1/c$ crystal, the
distortion groups the oxygen atoms into pairs. For a crystal at a
density of 9.8 \gcc, this implies every oxygen atom has one nearest
neighbor that is 1.45 \AA~away and two next-neighest neighbors at 1.49
\AA. There are five atoms at distances betwee 1.52 and 1.56 \AA. The
remaining four neighbors are between 1.70 and 1.73 \AA~away.

We performed two sets of Gibbs free energy calculations for the
$P2_1/c$ structure: one using the cell parameters derived from
ground-state relaxation, and the other refined with constant-pressure
MD simulations. Figure~\ref{fcc_P21c} compares the resulting Gibbs
free energies with those from a fcc crystal. Both sets of $P2_1/c$
calculations demonstrate clearly that there is a transition from fcc
to $P2_1/c$. The lattice parameter refinement lowers the Gibbs free
energy of the $P2_1/c$ structure slightly, which lowers the transition
pressure by about 1 Mbar to a value of 22.8$\pm$0.5 Mbar, which we
consider our most accurate prediction for the transition pressure from
fcc to the $P2_1/c$ structure. To obtain an estimate how sensitive
this prediction depends on the exchange-correlation functional, we we
recomputed the Gibbs free energy between the $P2_1/c$ and fcc
structures within the LDA. We obtained a slightly lower transition
pressure of 21.7 $\pm$ 0.3 Mbar. The magnitude of this difference is
not unexpected because LDA typically predicts higher densities and
lower transition pressures~\cite{driver-2010} than are obtained with
experiments and DFT calculations with other functionals. This was in
fact one of the original motivations for developing the PBE
functional~\cite{PBE}.

Figure~\ref{fcc_P21c} also shows the Gibbs free energy
difference between fcc and $P2_1/c$ split up into its three
components. The $P \Delta V$ term is negative and confirms that the
$P2_1/c$ structure is more dense than fcc at the same pressure and
temperature. Since fcc is a close packed structure as far as the
oxygen sub-lattice is concerned, this density change must be
accommodated by the hydrogen atoms or by changes in the electronic
structure. $-T\Delta S$ term suggests that significant changes occur
in the hydrogen sub-system. This term is positive and thus counteracts
a transformation to $P2_1/c$. It also reflects a reduction in the
mobility of the hydrogen atoms, which is consistent with the
non-uniform density of hydrogen atoms shown in Fig.~\ref{hcp_P21c}. We
consider this change to a less uniform hydrogen sub-system to be the
primary reason for a change from a close packed superionic system to
the $P2_1/c$ structure.

The difference in internal energy in Fig.~\ref{fcc_P21c} is
small compared to the other Gibbs free constituents but it decreases
with pressure and contributes to the energy balance that lead to a
transition pressure of 22.8 Mbar at 4000 K.

\begin{figure}[htb]
\includegraphics[width=140mm]{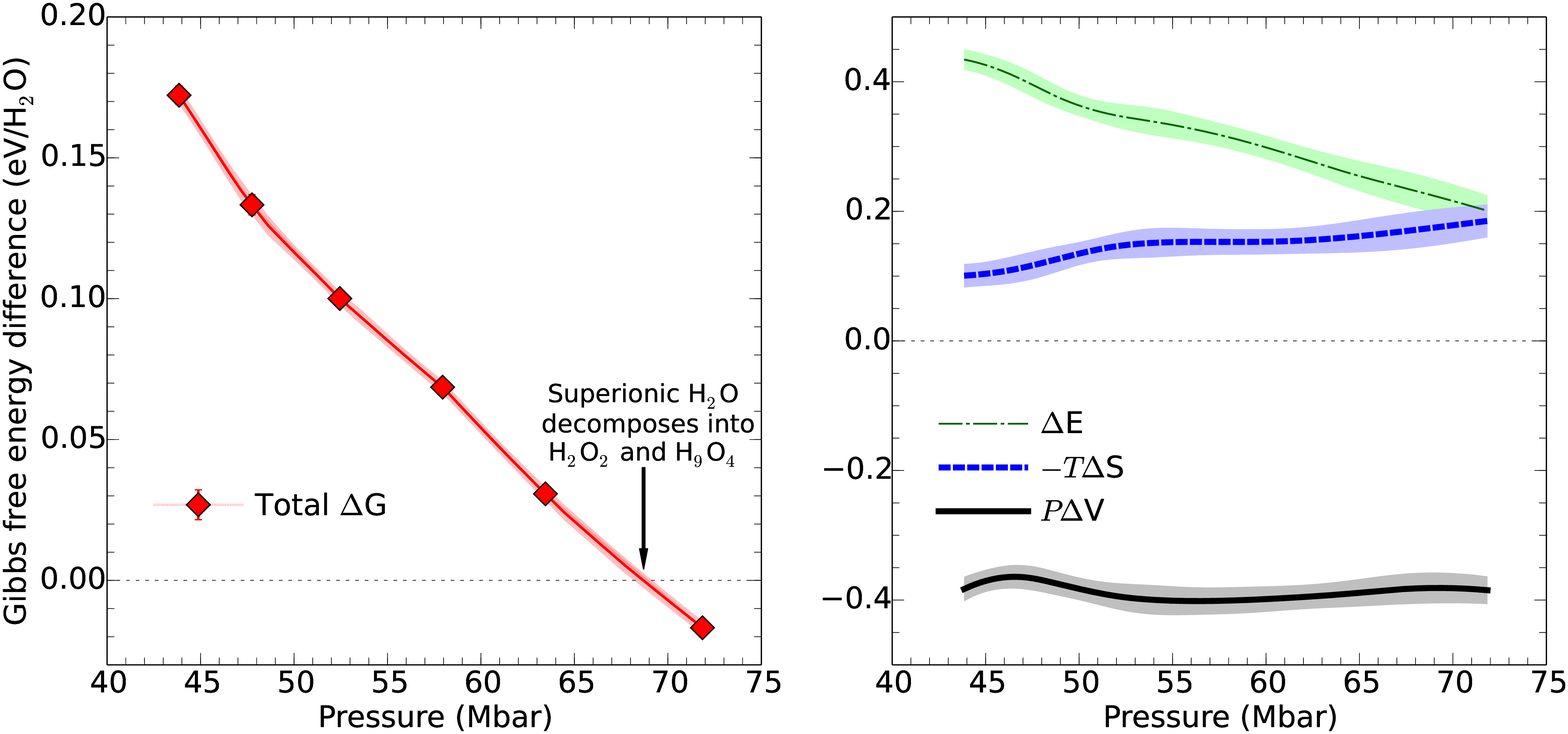}
\caption{Gibbs free energy difference, $\Delta G (\rm
  eV/H_2O) = [ G(H_2O_2) + 2 G(H_9O_4)]/10 - G(H_2O)$, for 5000 K
  as a function of pressure. The change of sign on the upper panel
  demonstrates the H$_2$O decomposes into H$_2$O$_2$ and H$_9$O$_4$ at
  the a pressure of 68.7 $\pm$ 0.5 Mbar. The right panel shows the
  different components on the Gibbs free energy difference.}
\label{G_H2O2}
\end{figure}

\section{DECOMPOSITION OF WATER INTO SUPERIONIC H$_2$O$_2$ AND H$_9$O$_4$}

\begin{figure}[htb]
\includegraphics[width=\smallerfigurewidth]{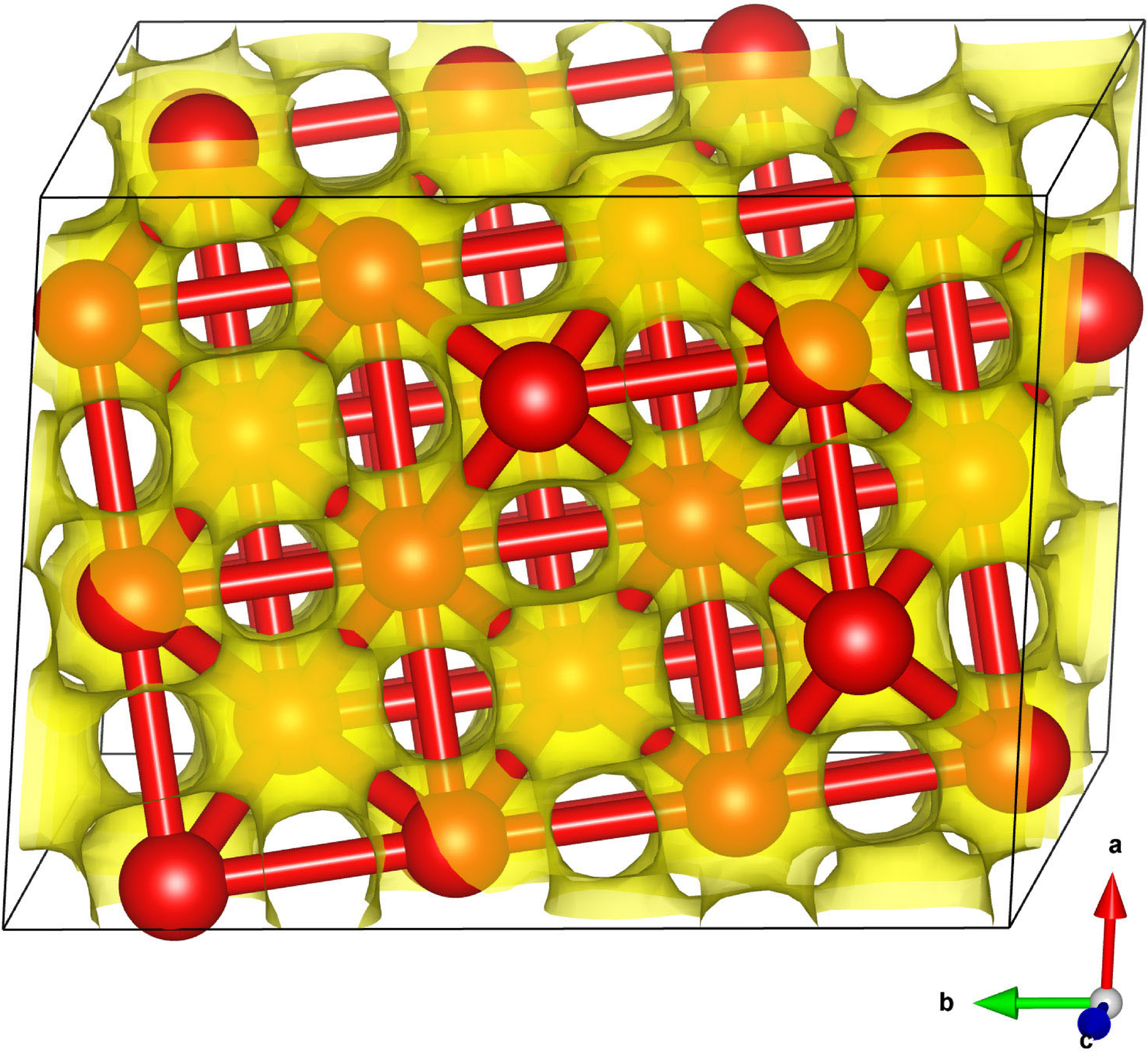}
\includegraphics[width=\figurewidth]{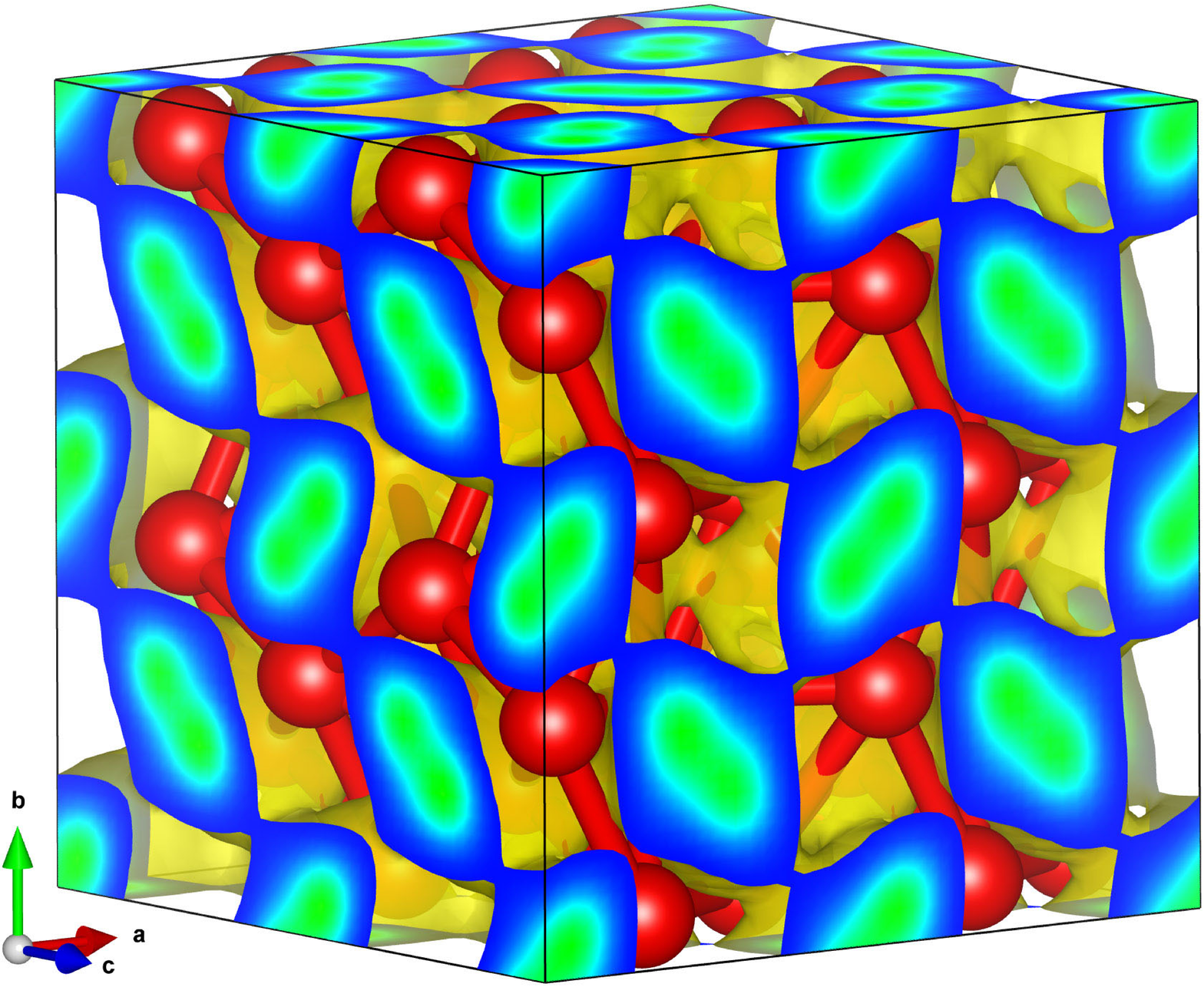}
\caption{Isosurfaces of the density of hydrogen atoms
  are compared for H$_9$O$_4$ (left) and H$_2$O$_2$ (right).}
\label{pictures_H2O2_H9O4}
\end{figure}

With a ground-state random structure search, Pickard {\it et
  al.}~\cite{Pickard2013} predicted water ice to decompose into
H$_2$O$_2$ and a hydrogen-rich structure like H$_9$O$_4$ at
approximately 50 Mbar. Since this exceeds the density range that we
have discussed so far, we now report simulation results from 10 $-$
13.5 \gcc. We performed DFT-MD simulations of H$_2$O$_2$ at constant-volume 
in a cubic (2~0~0,~0~2~0,~0~0~2) supercell with 32 formula
units. We started from the ground-state geometry at a density of 13.3
\gcc, which corresponds to a pressure of 50 Mbar, and gradually
increased the temperature during the MD simulation. At a temperature
of 4300$\,$K the hydrogen atoms become mobile while the oxygen atoms
remained confined to their lattice sites and the system assumed a
superionic state (Fig.~\ref{pictures_H2O2_H9O4}). This transition
temperature is confirmed by subsequent cooling simulations that showed
the system spontaneously transforming back to a solid state.

The oxygen sub-lattice appears to be significantly more stable than in
close-packed simulations at comparable conditions. In H$_2$O$_2$
simulations at 13.3 \gcc, 51.9 Mbar, and 6000 K, the root mean squared
displacement of the oxygen atoms from their $Pa \bar 3$ lattice sites
was only 0.14 \AA~while in simulations of fcc H$_2$O at comparable
$P$-$T$ conditions was approximately 0.21 \AA. This observation
prompted us to perform a number of additional investigations. First we
filled the superionic structure step by step with additional hydrogen
atoms. We were able to reach a H$_2$O stoichiometry while maintaining
a stable superionic structure. Any further addition of hydrogen atom
destabilized the $Pa \bar 3$ oxygen sub-lattice, however. We performed
Gibbs free energy calculations of this superionic H$_2$O structure at
5, 6, 8, 10 and 11 \gcc. Even though this structure exhibits a density
0.5\% higher than that of close-packed structures at same pressure,
its Gibbs free energy was always significantly higher (Fig.~\ref{GP}).
We therefore concluded the superionic water with a $Pa \bar 3$ oxygen
sub-lattice is not thermodynamically stable.

Since the $Pa \bar 3$ oxygen sub-lattice exhibits such an efficient
packing, we also investigated whether it would lead to H$_2$O
groundstate structures that are more stable than existing
predictions~\cite{McMahon2011,Ji2011,WangMa2011,Hermann2011,Pickard2013}.
We thus performed a ground-state random structure search for favorable
hydrogen positions in the H$_2$O stoichiometry while starting from a
$Pa \bar 3$ oxygen sub-lattice each time. As most stable, we
identified another structure with $P2_1/c$ symmetry that differed from
that of Ji \etal \cite{Ji2011}. Still, this $P2_1/c$ structure has a
higher enthalpy than other proposed ground-state structures for all
densities under consideration.  Therefore, we did not study
hydrogen-filled $Pa \bar 3$ oxygen sub-lattices any further.

For simulations of superionic H$_9$O$_4$, we constructed
(2~1~-1, -2~2~1, -2~0~-1) supercells with 156 atoms (12 formula units)
starting from the the ground-state geometries~\cite{Pickard2013}. This
supercell was chosen to be comparable in size to our other
simulations. During our heating simulations, the H$_9$O$_4$ system
assumed a superionic state (Fig.~\ref{pictures_H2O2_H9O4}) at a
temperature of approximately 1600 K.

We performed Gibbs free energy calculations for H$_2$O, H$_2$O$_2$,
and H$_9$O$_4$ structures at 5000 K for a series of densities in order
to cover the pressure interval from 45 to 70 Mbar. To demonstrate
that H$_2$O decomposes into H$_2$O$_2$ and H$_9$O$_4$, we need to show
that there is a sign change in Gibbs free energy difference per H$_2$O formula unit,
\begin{equation}
\Delta G = [ \, G({\rm H}_2{\rm O}_2) + 2 \, G({\rm H}_9{\rm O}_4) \,] /10  - G({\rm H}_2{\rm O})\;.
\end{equation}
Figure~\ref{G_H2O2} shows that such a sign change indeed occurs at a
pressure of 68.7 $\pm$ 0.5 Mbar implying that superionic H$_2$O
decomposed into a heterogeneous mixture of superionic H$_2$O$_2$,and
H$_9$O$_4$ at this pressure. In the lower panel of Fig.~\ref{G_H2O2},
we have separated the Gibbs free energy difference into its different
components. The $P \Delta V$ term is consistently large and negative,
implying that the decomposition of superionic H$_2$O is primarily
triggered by a more efficient packing in the H$_2$O$_2$ and H$_9$O$_4$
crystal structures. Consistent with this argument, we find that sign
of the $- T\Delta S$ term is positive, implying the hydrogen atoms
appear to have slightly more space to move around in the H$_2$O
structure. The $\Delta E$ term is positive and favors an H$_2$O
stoichiometry but it decays with increasing pressure. This trend
eventually changes the sign of the Gibbs free energy balance, which
leads to the decomposition of H$_2$O at 68.7 $\pm$ 0.5 Mbar. This
prediction is at variance with the recent work by French {\it et
  al.}~\cite{French2016} who suggested that the fcc superionic phase
transforms back to bcc at 50$\pm$10 Mbar at 5000 K. In their work,
only bcc and fcc phases were considered and the entropy was derived
from the vibrational density of states (VDOS). As stated in
Ref.~\cite{French2016}, this method is less accurate than the TDI
technique that we employed here for the following reasons. There are
different contributions to the entropy of many-body systems that need
to treated separately. French {\it et al.} employ a two-phase model to
split the VDOS in gas-like and solid-like parts. This splitting and
the treatment of each term rely to some degree on approximations. In
comparison, the TDI technique is simpler. As long as Eq.~\ref{tdint}
is evaluated with sufficient accuracy, all nuclear and electronic
contributions to the entropy are included~\cite{Militzer2013}. The
VDOS approach does, however, allow one to include nuclear quantum
effects, which we do not consider here.

  Using the TDI approach, we determined a decomposition pressure of
68.7 $\pm$ 0.5 Mbar for superionic water that is higher than the 50
Mbar that was predicted for the decomposition of solid H$_2$O ice at
zero temperature~\cite{Pickard2013}. Such a pressure shift is not
unexpected since we are dealing with different oxygen sub-lattices
and, more importantly, an entropy term, $-T\Delta S$, which is only
relevant at finite temperature. This term is positive and
approximately 0.1 eV/H$_2$O. Fig.~\ref{G_H2O2} shows that such a Gibbs
free energy change shift the transition pressure by +16 Mbar. This
lets us conclude that the entropy associated with the hydrogen motion
is the primary reason why our decomposition pressure for superionic
water is higher that for water ice.

\section{CONCLUSIONS}

We have presented more information about the phase diagram of water at
megabar pressures by performing {\it ab initio} simulations of
superionic water for a series of candidate structures. We showed that
the several close packed structures, fcc, hcp, and dhcp, have very
similar Gibbs free energies. The differences are on the order of
$\sim$0.02 eV/H$_2$O. Given the statistical and finite size
uncertainties of our current Gibbs free energy calculations, we are
not able to predict with certainty which superionic structure will be
assumed in the pressure interval from approximately 1.0 to 23 Mbar. It
is possible, however, that several stable phases may be realized or
possibly co-exist depending on pressure, temperature, and types of the
hydrogen isotope. For higher pressures, we find superionic water to
transform into novel structure with $P2_1/c$ symmetry, in which
hydrogen motion is more restricted.  This structure may be obtained
through a distortion of an hcp crystal.

At a much higher pressure of 68.7 $\pm$ 0.5 Mbar and temperature of
5000 K, we predict superionic water to decompose into superionic
H$_2$O$_2$ and superionic H$_9$O$_4$ structures. While an analysis of
the temperature dependence of this transition is beyond the scope of
this work, the predicted decomposition is consistent with the
decomposition that was proposed for water ice at 50 Mbar but the
superionic transition pressure is a bit higher because of the entropic
contribution from the hydrogen atoms to the Gibbs free energy balance.

\subsection{ACKNOWLEDGMENTS}
This work was supported from NSF and NASA.  Computational resources at
NCCS were used.  Part of the work by S.Z. was performed under the
auspices of the U.S. Department of Energy by Lawrence Livermore
National Laboratory under Contract No. DE-AC52-07NA27344.  We thank K.
Driver for comments.



\begin{thebibliography}{47}
\expandafter\ifx\csname natexlab\endcsname\relax\def\natexlab#1{#1}\fi
\expandafter\ifx\csname bibnamefont\endcsname\relax
  \def\bibnamefont#1{#1}\fi
\expandafter\ifx\csname bibfnamefont\endcsname\relax
  \def\bibfnamefont#1{#1}\fi
\expandafter\ifx\csname citenamefont\endcsname\relax
  \def\citenamefont#1{#1}\fi
\expandafter\ifx\csname url\endcsname\relax
  \def\url#1{\texttt{#1}}\fi
\expandafter\ifx\csname urlprefix\endcsname\relax\def\urlprefix{URL }\fi
\providecommand{\bibinfo}[2]{#2}
\providecommand{\eprint}[2][]{\url{#2}}

\bibitem[{\citenamefont{Hubbard}(1984)}]{hubbard_planets}
\bibinfo{author}{\bibfnamefont{W.~B.} \bibnamefont{Hubbard}},
  \emph{\bibinfo{title}{Planetary Interiors}} (\bibinfo{publisher}{University
  of Arizona Press}, \bibinfo{address}{Tucson, AZ}, \bibinfo{year}{1984}).

\bibitem[{\citenamefont{Ness et~al.}(1986)\citenamefont{Ness, Acuna, Behannon,
  Burlaga, Connery, Lepping, and Neubauer}}]{Ness1986}
\bibinfo{author}{\bibfnamefont{N.~F.} \bibnamefont{Ness}},
  \bibinfo{author}{\bibfnamefont{M.~H.} \bibnamefont{Acuna}},
  \bibinfo{author}{\bibfnamefont{K.~W.} \bibnamefont{Behannon}},
  \bibinfo{author}{\bibfnamefont{L.~F.} \bibnamefont{Burlaga}},
  \bibinfo{author}{\bibfnamefont{J.~E.~P.} \bibnamefont{Connery}},
  \bibinfo{author}{\bibfnamefont{R.~P.} \bibnamefont{Lepping}},
  \bibnamefont{and} \bibinfo{author}{\bibfnamefont{F.~M.}
  \bibnamefont{Neubauer}}, \bibinfo{journal}{Science}
  \textbf{\bibinfo{volume}{233}}, \bibinfo{pages}{85} (\bibinfo{year}{1986}).

\bibitem[{\citenamefont{Stanley and Bloxham}(2004)}]{Stanley2004}
\bibinfo{author}{\bibfnamefont{S.}~\bibnamefont{Stanley}} \bibnamefont{and}
  \bibinfo{author}{\bibfnamefont{J.}~\bibnamefont{Bloxham}},
  \bibinfo{journal}{Nature} \textbf{\bibinfo{volume}{428}},
  \bibinfo{pages}{151} (\bibinfo{year}{2004}).

\bibitem[{\citenamefont{Cavazzoni et~al.}(1999)\citenamefont{Cavazzoni,
  Chiarotti, Scandolo, Tosatti, Bernasconi, and Parrinello}}]{cavazzoni}
\bibinfo{author}{\bibfnamefont{C.}~\bibnamefont{Cavazzoni}},
  \bibinfo{author}{\bibfnamefont{G.~L.} \bibnamefont{Chiarotti}},
  \bibinfo{author}{\bibfnamefont{S.}~\bibnamefont{Scandolo}},
  \bibinfo{author}{\bibfnamefont{E.}~\bibnamefont{Tosatti}},
  \bibinfo{author}{\bibfnamefont{M.}~\bibnamefont{Bernasconi}},
  \bibnamefont{and}
  \bibinfo{author}{\bibfnamefont{M.}~\bibnamefont{Parrinello}},
  \bibinfo{journal}{Nature} \textbf{\bibinfo{volume}{283}}, \bibinfo{pages}{44}
  (\bibinfo{year}{1999}).

\bibitem[{\citenamefont{Goldman et~al.}(2005)\citenamefont{Goldman, Fried, Kuo,
  and Mundy}}]{goldman2005}
\bibinfo{author}{\bibfnamefont{N.}~\bibnamefont{Goldman}},
  \bibinfo{author}{\bibfnamefont{L.~E.} \bibnamefont{Fried}},
  \bibinfo{author}{\bibfnamefont{I.-.~W.} \bibnamefont{Kuo}}, \bibnamefont{and}
  \bibinfo{author}{\bibfnamefont{C.~J.} \bibnamefont{Mundy}},
  \bibinfo{journal}{Phys. Rev. Lett.} \textbf{\bibinfo{volume}{94}},
  \bibinfo{pages}{217801} (\bibinfo{year}{2005}).

\bibitem[{\citenamefont{Mattsson and Desjarlais}(2006)}]{mattson2006}
\bibinfo{author}{\bibfnamefont{T.~R.} \bibnamefont{Mattsson}} \bibnamefont{and}
  \bibinfo{author}{\bibfnamefont{M.~P.} \bibnamefont{Desjarlais}},
  \bibinfo{journal}{Phys. Rev. Lett.} \textbf{\bibinfo{volume}{97}},
  \bibinfo{pages}{017801} (\bibinfo{year}{2006}).

\bibitem[{\citenamefont{French et~al.}(2009)\citenamefont{French, Mattsson,
  Nettelmann, and Redmer}}]{french-prb-09}
\bibinfo{author}{\bibfnamefont{M.}~\bibnamefont{French}},
  \bibinfo{author}{\bibfnamefont{T.~R.} \bibnamefont{Mattsson}},
  \bibinfo{author}{\bibfnamefont{N.}~\bibnamefont{Nettelmann}},
  \bibnamefont{and} \bibinfo{author}{\bibfnamefont{R.}~\bibnamefont{Redmer}},
  \bibinfo{journal}{Phys. Rev. B} \textbf{\bibinfo{volume}{79}},
  \bibinfo{pages}{054107} (\bibinfo{year}{2009}).

\bibitem[{\citenamefont{Wilson et~al.}(2013)\citenamefont{Wilson, Wong, and
  Militzer}}]{WilsonWongMilitzer2013}
\bibinfo{author}{\bibfnamefont{H.~F.} \bibnamefont{Wilson}},
  \bibinfo{author}{\bibfnamefont{M.~L.} \bibnamefont{Wong}}, \bibnamefont{and}
  \bibinfo{author}{\bibfnamefont{B.}~\bibnamefont{Militzer}},
  \bibinfo{journal}{Phys. Rev. Lett.} \textbf{\bibinfo{volume}{110}},
  \bibinfo{pages}{151102} (\bibinfo{year}{2013}).

\bibitem[{\citenamefont{Hernandez and Caracas}(2016)}]{caracas2016}
\bibinfo{author}{\bibfnamefont{J.-A.} \bibnamefont{Hernandez}}
  \bibnamefont{and} \bibinfo{author}{\bibfnamefont{R.}~\bibnamefont{Caracas}},
  \bibinfo{journal}{Phys. Rev. Lett.} \textbf{\bibinfo{volume}{117}},
  \bibinfo{pages}{135503} (\bibinfo{year}{2016}).

\bibitem[{gon()}]{goncharov2009}
\bibinfo{note}{A.~F.~Goncharov~{\it et~al.}, {\it J. Chem. Phys.} {\bf 130},
  124514 (2009).}

\bibitem[{hem()}]{hemley-ice-1987}
\bibinfo{note}{R.~J.~Hemley~{\it et~al.}, {\it Nature} {\bf 330}, 737 (1987).}

\bibitem[{Gon()}]{Goncharov1996}
\bibinfo{note}{A.~F.~Goncharov~{\it et~al.}, {\it Science} {\bf 273}, 218
  (1996).}

\bibitem[{\citenamefont{Loubeyre et~al.}(1999)\citenamefont{Loubeyre,
  LeToullec, Wolanin, Han, and Hausermann}}]{Loubeyre1999}
\bibinfo{author}{\bibfnamefont{P.}~\bibnamefont{Loubeyre}},
  \bibinfo{author}{\bibfnamefont{R.}~\bibnamefont{LeToullec}},
  \bibinfo{author}{\bibfnamefont{E.}~\bibnamefont{Wolanin}},
  \bibinfo{author}{\bibfnamefont{M.}~\bibnamefont{Han}}, \bibnamefont{and}
  \bibinfo{author}{\bibfnamefont{D.}~\bibnamefont{Hausermann}},
  \bibinfo{journal}{Nature} \textbf{\bibinfo{volume}{397}},
  \bibinfo{pages}{503} (\bibinfo{year}{1999}).

\bibitem[{lee()}]{lee2006_short}
\bibinfo{note}{K. K. M. Lee {\it et al.} {\it J. Chem. Phys.} {\bf 125}, 014701
  (2006).}

\bibitem[{\citenamefont{Knudson et~al.}(2012)\citenamefont{Knudson, Desjarlais,
  Lemke, Mattsson, French, Nettelmann, and Redmer}}]{KnudsonWater2012}
\bibinfo{author}{\bibfnamefont{M.~D.} \bibnamefont{Knudson}},
  \bibinfo{author}{\bibfnamefont{M.~P.} \bibnamefont{Desjarlais}},
  \bibinfo{author}{\bibfnamefont{R.~W.} \bibnamefont{Lemke}},
  \bibinfo{author}{\bibfnamefont{T.~R.} \bibnamefont{Mattsson}},
  \bibinfo{author}{\bibfnamefont{M.}~\bibnamefont{French}},
  \bibinfo{author}{\bibfnamefont{N.}~\bibnamefont{Nettelmann}},
  \bibnamefont{and} \bibinfo{author}{\bibfnamefont{R.}~\bibnamefont{Redmer}},
  \bibinfo{journal}{Phys. Rev. Lett.} \textbf{\bibinfo{volume}{108}},
  \bibinfo{pages}{091102} (\bibinfo{year}{2012}).

\bibitem[{\citenamefont{Wang}(2013)}]{WangRamp2013}
\bibinfo{author}{\bibfnamefont{J.}~\bibnamefont{Wang {\it et al.}}}, \bibinfo{journal}{J.
  Appl. Phys.} \textbf{\bibinfo{volume}{114}}, \bibinfo{pages}{023513}
  (\bibinfo{year}{2013}).

\bibitem[{\citenamefont{Militzer and Wilson}(2010)}]{MilitzerWilson2010}
\bibinfo{author}{\bibfnamefont{B.}~\bibnamefont{Militzer}} \bibnamefont{and}
  \bibinfo{author}{\bibfnamefont{H.~F.} \bibnamefont{Wilson}},
  \bibinfo{journal}{Phys. Rev. Lett.} \textbf{\bibinfo{volume}{105}},
  \bibinfo{pages}{195701} (\bibinfo{year}{2010}).

\bibitem[{\citenamefont{McMahon}(2011)}]{McMahon2011}
\bibinfo{author}{\bibfnamefont{J.~F.} \bibnamefont{McMahon}},
  \bibinfo{journal}{Phys. Rev. B} \textbf{\bibinfo{volume}{84}},
  \bibinfo{pages}{220104(R)} (\bibinfo{year}{2011}).

\bibitem[{\citenamefont{Ji et~al.}(2011)\citenamefont{Ji, Umemoto, Wang, Ho,
  and M.Wentzcovitch}}]{Ji2011}
\bibinfo{author}{\bibfnamefont{M.}~\bibnamefont{Ji}},
  \bibinfo{author}{\bibfnamefont{K.}~\bibnamefont{Umemoto}},
  \bibinfo{author}{\bibfnamefont{C.-Z.} \bibnamefont{Wang}},
  \bibinfo{author}{\bibfnamefont{K.-M.} \bibnamefont{Ho}}, \bibnamefont{and}
  \bibinfo{author}{\bibfnamefont{R.}~\bibnamefont{M.Wentzcovitch}},
  \bibinfo{journal}{Phys. Rev. B} \textbf{\bibinfo{volume}{84}},
  \bibinfo{pages}{220105(R)} (\bibinfo{year}{2011}).

\bibitem[{\citenamefont{Wang et~al.}(2011)\citenamefont{Wang, Liu, Lv, Zhu,
  Wang, and Ma}}]{WangMa2011}
\bibinfo{author}{\bibfnamefont{Y.}~\bibnamefont{Wang}},
  \bibinfo{author}{\bibfnamefont{H.}~\bibnamefont{Liu}},
  \bibinfo{author}{\bibfnamefont{J.}~\bibnamefont{Lv}},
  \bibinfo{author}{\bibfnamefont{L.}~\bibnamefont{Zhu}},
  \bibinfo{author}{\bibfnamefont{H.}~\bibnamefont{Wang}}, \bibnamefont{and}
  \bibinfo{author}{\bibfnamefont{Y.}~\bibnamefont{Ma}},
  \bibinfo{journal}{Nature Comm.} \textbf{\bibinfo{volume}{2}},
  \bibinfo{pages}{563} (\bibinfo{year}{2011}).

\bibitem[{\citenamefont{Hermann et~al.}(2011)\citenamefont{Hermann, Ashcroft,
  and Hoffmann}}]{Hermann2011}
\bibinfo{author}{\bibfnamefont{A.}~\bibnamefont{Hermann}},
  \bibinfo{author}{\bibfnamefont{N.~W.} \bibnamefont{Ashcroft}},
  \bibnamefont{and} \bibinfo{author}{\bibfnamefont{R.}~\bibnamefont{Hoffmann}},
  \bibinfo{journal}{Proc. Nat. Acad. Sci.} \textbf{\bibinfo{volume}{109}},
  \bibinfo{pages}{745} (\bibinfo{year}{2011}).

\bibitem[{\citenamefont{Zhang et~al.}(2013)\citenamefont{Zhang, Wilson, Driver,
  and Militzer}}]{H4O}
\bibinfo{author}{\bibfnamefont{S.}~\bibnamefont{Zhang}},
  \bibinfo{author}{\bibfnamefont{H.~F.} \bibnamefont{Wilson}},
  \bibinfo{author}{\bibfnamefont{K.~P.} \bibnamefont{Driver}},
  \bibnamefont{and} \bibinfo{author}{\bibfnamefont{B.}~\bibnamefont{Militzer}},
  \bibinfo{journal}{Phys. Rev. B} \textbf{\bibinfo{volume}{87}},
  \bibinfo{pages}{024112} (\bibinfo{year}{2013}).

\bibitem[{\citenamefont{Pickard et~al.}(2013)\citenamefont{Pickard,
  Martinez-Canales, and Needs}}]{Pickard2013}
\bibinfo{author}{\bibfnamefont{C.~J.} \bibnamefont{Pickard}},
  \bibinfo{author}{\bibfnamefont{M.}~\bibnamefont{Martinez-Canales}},
  \bibnamefont{and} \bibinfo{author}{\bibfnamefont{R.~J.} \bibnamefont{Needs}},
  \bibinfo{journal}{Phys. Rev. Lett.} \textbf{\bibinfo{volume}{110}},
  \bibinfo{pages}{245701} (\bibinfo{year}{2013}).

\bibitem[{\citenamefont{Sun et~al.}(2015)\citenamefont{Sun, Clark, Torquato,
  and Car}}]{Princeton}
\bibinfo{author}{\bibfnamefont{J.}~\bibnamefont{Sun}},
  \bibinfo{author}{\bibfnamefont{B.~K.} \bibnamefont{Clark}},
  \bibinfo{author}{\bibfnamefont{S.}~\bibnamefont{Torquato}}, \bibnamefont{and}
  \bibinfo{author}{\bibfnamefont{R.}~\bibnamefont{Car}},
  \bibinfo{journal}{Nature Commun.} \textbf{\bibinfo{volume}{6}},
  \bibinfo{pages}{8156} (\bibinfo{year}{2015}).

\bibitem[{APS()}]{APS}
\bibinfo{note}{In the same session of American Physical Society meeting in
  March 2014 where we presented the majority of the results from this work, J.
  Sun, B. Clark, and R. Car also predicted a phase transition in superionic
  H$_2$O water based on spontaneous transitions in constant pressure
  simulations of $P2_1$ supercells.}

\bibitem[{vas()}]{vasp}
\bibinfo{note}{G. Kresse and J. Hafner, Phys. Rev. B 47, 558 (1993); G. Kresse
  and J. Hafner, Phys. Rev. B 49, 14251 (1994); G. Kresse and J. Furthm\"uller,
  Comput. Mat. Sci. 6, 15 (1996); G. Kresse and J. Furthm\"uller, Phys. Rev. B
  54, 11169 (1996).}

\bibitem[{\citenamefont{Bl\"ochl}(1994)}]{paw}
\bibinfo{author}{\bibfnamefont{P.~E.} \bibnamefont{Bl\"ochl}},
  \bibinfo{journal}{Phys. Rev. B} \textbf{\bibinfo{volume}{50}},
  \bibinfo{pages}{17953} (\bibinfo{year}{1994}).

\bibitem[{\citenamefont{Perdew et~al.}(1996)\citenamefont{Perdew, Burke, and
  Ernzerhof}}]{PBE}
\bibinfo{author}{\bibfnamefont{J.~P.} \bibnamefont{Perdew}},
  \bibinfo{author}{\bibfnamefont{K.}~\bibnamefont{Burke}}, \bibnamefont{and}
  \bibinfo{author}{\bibfnamefont{M.}~\bibnamefont{Ernzerhof}},
  \bibinfo{journal}{Phys. Rev. Lett.} \textbf{\bibinfo{volume}{77}},
  \bibinfo{pages}{3865} (\bibinfo{year}{1996}).

\bibitem[{\citenamefont{Monkhorst and Pack}(1976)}]{MP76}
\bibinfo{author}{\bibfnamefont{H.}~\bibnamefont{Monkhorst}} \bibnamefont{and}
  \bibinfo{author}{\bibfnamefont{J.}~\bibnamefont{Pack}},
  \bibinfo{journal}{Phys. Rev. B.} \textbf{\bibinfo{volume}{13}},
  \bibinfo{pages}{5188} (\bibinfo{year}{1976}).

\bibitem[{\citenamefont{Mermin}(1965)}]{mermin}
\bibinfo{author}{\bibfnamefont{N.~D.} \bibnamefont{Mermin}},
  \bibinfo{journal}{Phys. Rev.} \textbf{\bibinfo{volume}{137}},
  \bibinfo{pages}{A1441} (\bibinfo{year}{1965}).

\bibitem[{\citenamefont{Militzer}(2016)}]{supercells}
\bibinfo{author}{\bibfnamefont{B.}~\bibnamefont{Militzer}},
  \bibinfo{journal}{J. High Energy Density Phys.}
  \textbf{\bibinfo{volume}{21}}, \bibinfo{pages}{8} (\bibinfo{year}{2016}).

\bibitem[{\citenamefont{Hernandez}(2001)}]{Hernandez2001}
\bibinfo{author}{\bibfnamefont{E.}~\bibnamefont{Hernandez}},
  \bibinfo{journal}{J. Chem. Phys.} \textbf{\bibinfo{volume}{115}},
  \bibinfo{pages}{10282} (\bibinfo{year}{2001}).

\bibitem[{\citenamefont{de~Wijs et~al.}(1998)\citenamefont{de~Wijs, Kresse, and
  Gillan}}]{Wijs1998}
\bibinfo{author}{\bibfnamefont{G.~A.} \bibnamefont{de~Wijs}},
  \bibinfo{author}{\bibfnamefont{G.}~\bibnamefont{Kresse}}, \bibnamefont{and}
  \bibinfo{author}{\bibfnamefont{M.~J.} \bibnamefont{Gillan}},
  \bibinfo{journal}{Phys. Rev. B} \textbf{\bibinfo{volume}{57}},
  \bibinfo{pages}{8223} (\bibinfo{year}{1998}).

\bibitem[{\citenamefont{Morales et~al.}(2009)\citenamefont{Morales, Pierleoni,
  Schwegler, and Ceperley}}]{Morales2009}
\bibinfo{author}{\bibfnamefont{M.~A.} \bibnamefont{Morales}},
  \bibinfo{author}{\bibfnamefont{C.}~\bibnamefont{Pierleoni}},
  \bibinfo{author}{\bibfnamefont{E.}~\bibnamefont{Schwegler}},
  \bibnamefont{and} \bibinfo{author}{\bibfnamefont{D.~M.}
  \bibnamefont{Ceperley}}, \bibinfo{journal}{Proc. Nat. Acad. Sci.}
  \textbf{\bibinfo{volume}{106}}, \bibinfo{pages}{1324} (\bibinfo{year}{2009}).

\bibitem[{\citenamefont{Wilson and Militzer}(2010)}]{WilsonMilitzer2010}
\bibinfo{author}{\bibfnamefont{H.~F.} \bibnamefont{Wilson}} \bibnamefont{and}
  \bibinfo{author}{\bibfnamefont{B.}~\bibnamefont{Militzer}},
  \bibinfo{journal}{Phys. Rev. Lett.} \textbf{\bibinfo{volume}{104}},
  \bibinfo{pages}{121101} (\bibinfo{year}{2010}).

\bibitem[{\citenamefont{Wilson and
  Militzer}(2012{\natexlab{a}})}]{WilsonMilitzer2012}
\bibinfo{author}{\bibfnamefont{H.~F.} \bibnamefont{Wilson}} \bibnamefont{and}
  \bibinfo{author}{\bibfnamefont{B.}~\bibnamefont{Militzer}},
  \bibinfo{journal}{Astrophys. J.} \textbf{\bibinfo{volume}{745}},
  \bibinfo{pages}{54} (\bibinfo{year}{2012}{\natexlab{a}}).

\bibitem[{\citenamefont{Wilson and
  Militzer}(2012{\natexlab{b}})}]{WilsonMilitzer2012b}
\bibinfo{author}{\bibfnamefont{H.~F.} \bibnamefont{Wilson}} \bibnamefont{and}
  \bibinfo{author}{\bibfnamefont{B.}~\bibnamefont{Militzer}},
  \bibinfo{journal}{Phys. Rev. Lett.} \textbf{\bibinfo{volume}{108}},
  \bibinfo{pages}{111101} (\bibinfo{year}{2012}{\natexlab{b}}).

\bibitem[{\citenamefont{Militzer}(2013)}]{Militzer2013}
\bibinfo{author}{\bibfnamefont{B.}~\bibnamefont{Militzer}},
  \bibinfo{journal}{Phys. Rev. B} \textbf{\bibinfo{volume}{87}},
  \bibinfo{pages}{014202} (\bibinfo{year}{2013}).

\bibitem[{\citenamefont{Wahl et~al.}(2013)\citenamefont{Wahl, Wilson, and
  Militzer}}]{Wahl2013}
\bibinfo{author}{\bibfnamefont{S.~M.} \bibnamefont{Wahl}},
  \bibinfo{author}{\bibfnamefont{H.~F.} \bibnamefont{Wilson}},
  \bibnamefont{and} \bibinfo{author}{\bibfnamefont{B.}~\bibnamefont{Militzer}},
  \bibinfo{journal}{Astrophys. J.} \textbf{\bibinfo{volume}{773}},
  \bibinfo{pages}{95} (\bibinfo{year}{2013}).

\bibitem[{\citenamefont{Gonzalez-Cataldo
  et~al.}(2014)\citenamefont{Gonzalez-Cataldo, Wilson, and
  Militzer}}]{Gonzalez2015}
\bibinfo{author}{\bibfnamefont{F.}~\bibnamefont{Gonzalez-Cataldo}},
  \bibinfo{author}{\bibfnamefont{H.~F.} \bibnamefont{Wilson}},
  \bibnamefont{and} \bibinfo{author}{\bibfnamefont{B.}~\bibnamefont{Militzer}},
  \bibinfo{journal}{Astrophys. J.} \textbf{\bibinfo{volume}{787}},
  \bibinfo{pages}{79} (\bibinfo{year}{2014}).

\bibitem[{\citenamefont{Soubiran and Militzer}(2015)}]{Soubiran2015}
\bibinfo{author}{\bibfnamefont{F.}~\bibnamefont{Soubiran}} \bibnamefont{and}
  \bibinfo{author}{\bibfnamefont{B.}~\bibnamefont{Militzer}},
  \bibinfo{journal}{Astrophysical J.} \textbf{\bibinfo{volume}{806}},
  \bibinfo{pages}{228} (\bibinfo{year}{2015}).

\bibitem[{\citenamefont{Soubiran and Militzer}(2016)}]{Soubiran2016}
\bibinfo{author}{\bibfnamefont{F.}~\bibnamefont{Soubiran}} \bibnamefont{and}
  \bibinfo{author}{\bibfnamefont{B.}~\bibnamefont{Militzer}},
  \bibinfo{journal}{Astrophysical J.} \textbf{\bibinfo{volume}{829}},
  \bibinfo{pages}{14} (\bibinfo{year}{2016}).

\bibitem[{\citenamefont{Izvekov et~al.}(2003)\citenamefont{Izvekov, Parrinello,
  Burnham, and Voth}}]{forcematching}
\bibinfo{author}{\bibfnamefont{S.}~\bibnamefont{Izvekov}},
  \bibinfo{author}{\bibfnamefont{M.}~\bibnamefont{Parrinello}},
  \bibinfo{author}{\bibfnamefont{C.~J.} \bibnamefont{Burnham}},
  \bibnamefont{and} \bibinfo{author}{\bibfnamefont{G.~A.} \bibnamefont{Voth}},
  \bibinfo{journal}{J. Chem. Phys.} \textbf{\bibinfo{volume}{120}},
  \bibinfo{pages}{10896} (\bibinfo{year}{2003}).

\bibitem[{\citenamefont{Wahl and Militzer}(2015)}]{Wahl2015}
\bibinfo{author}{\bibfnamefont{S.}~\bibnamefont{Wahl}} \bibnamefont{and}
  \bibinfo{author}{\bibfnamefont{B.}~\bibnamefont{Militzer}},
  \bibinfo{journal}{Earth and Planetary Science Letters}
  \textbf{\bibinfo{volume}{410}}, \bibinfo{pages}{25} (\bibinfo{year}{2015}).

\bibitem[{\citenamefont{Ma et~al.}(2008)\citenamefont{Ma, Oganov, and
  Xie}}]{MaSodium2008}
\bibinfo{author}{\bibfnamefont{Y.}~\bibnamefont{Ma}},
  \bibinfo{author}{\bibfnamefont{A.~R.} \bibnamefont{Oganov}},
  \bibnamefont{and} \bibinfo{author}{\bibfnamefont{Y.}~\bibnamefont{Xie}},
  \bibinfo{journal}{Phys. Rev. B} \textbf{\bibinfo{volume}{78}},
  \bibinfo{pages}{014102} (\bibinfo{year}{2008}).

\bibitem[{\citenamefont{Driver et~al.}(2010)\citenamefont{Driver, Cohen, Wu,
  Militzer, Rios, Towler, Needs, and Wilkins}}]{driver-2010}
\bibinfo{author}{\bibfnamefont{K.~P.} \bibnamefont{Driver}},
  \bibinfo{author}{\bibfnamefont{R.~E.} \bibnamefont{Cohen}},
  \bibinfo{author}{\bibfnamefont{Z.}~\bibnamefont{Wu}},
  \bibinfo{author}{\bibfnamefont{B.}~\bibnamefont{Militzer}},
  \bibinfo{author}{\bibfnamefont{P.~L.} \bibnamefont{Rios}},
  \bibinfo{author}{\bibfnamefont{M.~D.} \bibnamefont{Towler}},
  \bibinfo{author}{\bibfnamefont{R.~J.} \bibnamefont{Needs}}, \bibnamefont{and}
  \bibinfo{author}{\bibfnamefont{J.~W.} \bibnamefont{Wilkins}},
  \bibinfo{journal}{Proc. Nat. Acad. Sci.} \textbf{\bibinfo{volume}{107}},
  \bibinfo{pages}{9519} (\bibinfo{year}{2010}).

\bibitem[{\citenamefont{French et~al.}(2016)\citenamefont{French, Desjarlais,
  and Redmer}}]{French2016}
\bibinfo{author}{\bibfnamefont{M.}~\bibnamefont{French}},
  \bibinfo{author}{\bibfnamefont{M.~P.} \bibnamefont{Desjarlais}},
  \bibnamefont{and} \bibinfo{author}{\bibfnamefont{R.}~\bibnamefont{Redmer}},
  \bibinfo{journal}{Phys. Rev. E} \textbf{\bibinfo{volume}{93}},
  \bibinfo{pages}{022140} (\bibinfo{year}{2016}).

\end{thebibliography}
\end{document}